\documentclass{svmult}
\usepackage{graphicx} 
\usepackage{url}
\def\apj{{ApJ\ }}                 
\def\apjl{{ApJ\ }}                
\def\apjs{{ApJS\ }}               
\def\aap{{A\&A\ }}                
\def\mnras{{MNRAS\ }}             
\def\lea{\mathrel{\raise .4ex\hbox{\rlap{$<$}\lower 1.2ex\hbox{$\sim$}}}}
\def\gea{\mathrel{\raise .4ex\hbox{\rlap{$>$}\lower 1.2ex\hbox{$\sim$}}}}

\newcommand{\hMpc}{\, h^{-1}\, \mathrm{Mpc}}
\newcommand{\mx}{{\mathbf x}}
\newcommand{\mk}{{\mathbf k}}
\newcommand{\mr}{{\mathbf r}}
\begin{document}

\title*{The Large Scale Structure in the Universe: From
Power-Laws to Acoustic Peaks}
\titlerunning{From Power-Laws to Acoustic Peaks}
\author{Vicent J. Mart\'{\i}nez}
\institute{Observatori Astron\`omic, Universitat de Val\`encia, \\
Edifici d'Instituts d'Investigaci\'o, Pol\'{\i}gon La Coma, 46980 Paterna, Val\`encia, Spain. \\
\url{martinez@uv.es}}
\maketitle

The most popular tools for analysing  the large scale distribution
of galaxies are second-order spatial statistics such as the
two-point correlation function or its Fourier transform, the power
spectrum. In this review, we explain how our knowledge of cosmic
structures, encapsulated by these statistical descriptors, has
evolved since their first use when applied on the early galaxy
catalogues to the present generation of wide and deep redshift
surveys.\footnote{Being the first editor of this volume gives me the
opportunity of updating this review taking into account the more
recent developments in the field. I have used this opportunity
trying to incorporating ll the most challenging discovery in the study of
the galaxy distribution: the detection of Baryon Acoustic
Oscillations.}

\section{Introduction \label{sec:vmintroduction}}

As the reader can learn from this volume, there are mainly two
astronomical observations that provide the most relevant
cosmological data needed to probe any cosmological model: the Cosmic
Microwave Background radiation and the Large Scale Structure of the
Universe. This review deals with the second of these cosmological
fossils.  The statistical analysis of galaxy clustering
\index{galaxy clustering} has been progressing in parallel with the
development of the observations of the galaxy distribution (for a
review see e.g. Jones et al. \cite{VMjonesRMP04}). Since the
pioneering works by Hubble, measuring the distribution of the number
counts of galaxies in telescope fields and finding a log-Gaussian
distribution \cite{VMhubble34}, many authors have described the best
available data at each moment making use of the then well
established statistical tools. For example, F. Zwicky
\cite{VMzwicky53} used the ratio of clumpiness, the quotient between
the variance of the number counts and the expected quantity for a
Poisson distribution.

The first map of the sky revealing convincing clustering of galaxies
was the Lick survey undertaken by Shane and Wirtanen \cite{VMLick67}
. While the catalogue was in progress, two different approaches to
its statistical description were developed: The Neyman-Scott
approach and  the Correlation Function school named in this way by
Bernard Jones \cite{VMjones01}.

Jerzy Neyman and Elisabeth Scott were the first to consider the
galaxy distribution as a realisation of a homogeneous random point
process \cite{VMneyman52}.  They formulated {\it a priori}
statistical models to describe the clustering of galaxies and later
they tried to fit the parameters of the model by comparing it with
observations. In this way, they modeled the distribution of galaxy
clusters as a random superposition of groups following what now is
known in spatial statistics as a Neyman-Scott process, i.e.,
\index{Neyman-Scott process} a Poisson cluster process constructed
in two steps: first, a homogeneous Poisson process is generated by
randomly distributing a set of centres (or parent points); second, a
cluster of daughter points is scattered around each of the parent
points, according to a given density function. This idea
\cite{VMNSS53, VMpeebles74}  is the basis of the recent halo model
\cite{VMhalo} that successfully describes the statistics of the
matter distribution in structures of different sizes at different
scales: at small scales the halo model assumes that the distribution
is dominated by the density profiles of the dark matter halos, and
therefore correlations come mainly from intra-halo pairs. The most
popular density profile is that of Navarro, Frenk and White
\cite{VMNFW97}.

\begin{figure}
\centering
\resizebox{\textwidth}{!}{\includegraphics*{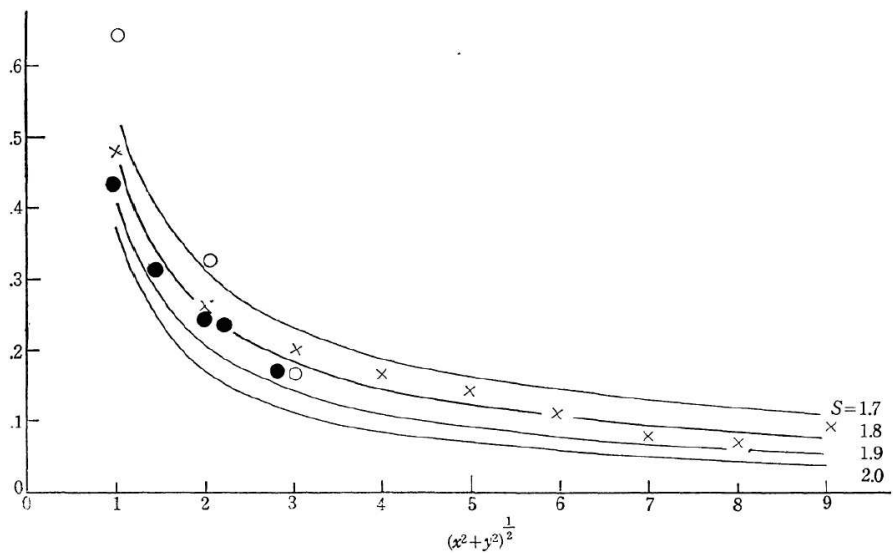}}
\caption{The first power-law fitting the spatial correlation
function of the distribution of galaxies after deprojecting from an
angular catalogue, reproduced from ~\cite{VMtotsuji69}. The filled
circles were obtained by Totsuji and Kihara, while the open circles
and crosses were derived by Neyman, Scott and Shane under the
assumption of their clustering model. The solid lines correspond to
power-law correlation functions $xi(r)=(r_0/r)^s$ with the value of
the exponent $s$ indicated in the legend.} \label{fig:totsuji}
\end{figure}

The second approach based on the correlation function was envisaged
first by Vera Rubin \cite{VMrubin54} and by D. Nelson Limber
\cite{VMlimber54}.  They thought that the galaxy distribution was in
fact a set of points sampled from an underlying continuous density
distribution that later was called the Poisson model by Peebles
\cite{VMpeebles80}. In   spatial statistics this is known as a Cox
process \cite{VMmartsaar}. \index{Cox process} They derived the
auto-correlation function from the variance of the number counts of
the on-going Lick survey. Moreover, Limber provided an integral
equation relating the angular and the spatial correlation function
valid for small angle separation (a special version of this equation
appears also in the paper by Rubin). The correlation function
\index{correlation!function} measures the clustering in excess
$[\xi(r)>0]$ or in defect $[\xi(r)<0]$ compared with a Poisson
distribution. It can be defined in terms of the probability $dP$ of
finding a galaxy in a small volume $dV$ lying at a distance $r$ of a
given galaxy
\begin{equation}
dP = n [1 + \xi(r)] dV.
\label{eq:xidef2}
\end{equation}
where $n$ is the mean number density over the whole sample volume
(see Section \ref{sec:vmcorrfunc} for a more formal definition.)
Totsuji and Kihara \cite{VMtotsuji69} were the first to obtain a
power-law \index{correlation!function!power-law} behaviour for the
spatial correlation function $\xi(r)=(r/r_0)^{-1.8}$ on the basis of
angular data taken from the Lick survey and making use of the Limber
equation. Moreover, as we can see in Fig. ~\ref{fig:totsuji}
reproduced from their paper, the observed correlation function of
the Lick survey is fitted to an early halo model -- the Neyman-Scott
process.

This remarkable power-law for the two-point correlation function has
dominated many of the analyses of the large scale structure for the
past three decades and more.

Complementary to the Lick catalog, other surveys mapped the large
scale distribution of clusters of galaxies, for example, the Palomar
Observatory Sky Survey was used by George Abell  to publish a
catalogue of 2,712 clusters of galaxies \cite{VMabell58} . Some of
them turned out not to be real clusters, but the majority were
genuine. Analyses of this and other samples of galaxy clusters have
yielded also power-law fits to the cluster-cluster correlation
function $\xi_{cc}(r)$ but with exponents and amplitudes varying in
a wider range, depending on selection effects, richness class, etc.
\cite{VMpostman92, VMnichol92, VMdalton94, VMpostman99,
VMborgani01}.

\section{Redshift Surveys \label{sec:vmredshift}}

Listing extragalactic objects and magnitudes as they appear
projected onto the celestial sphere was just the first step towards
obtaining a cartography of the universe. The second step was to
obtain distances by measuring redshifts using spectroscopy for a
large number of galaxies mapping large areas of the sky. This task
provided information about how the universe is structured now and in
the recent past.  In the eighties, the Center for Astrophysics
surveys played a leading role in the discovery of very large cosmic
structures in the distribution of the galaxies. The first ``slice of
the universe" \index{slice of the universe} compiled by de~Lapparent
et al. \cite{VMlapparent86} extended up to 150 $h^{-1}$ Mpc, a deep
distance at that time. The calculation of the correlation function
-- now in redshift space -- of the CfA catalogue confirmed the
power-law behaviour discovered by Totsuji and Kihara fourteen years
before \cite{VMDP83}. It is worth to mention however that redshift
distortions affect severely the correlation function at small
separations and a distinction between redshift and real space became
necessary.

\begin{figure}
\centering
\resizebox{\textwidth}{!}{\includegraphics*{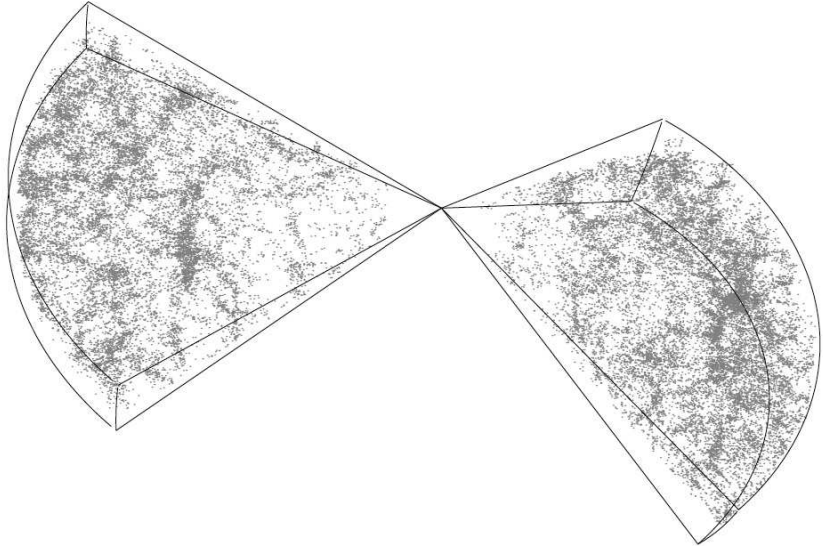}}
\caption{The two slices that conform the 2dfGRS showing the galaxy
distribution up to a distance of $400 \hMpc$. The left slice lies in
the direction close to the North Galactic Pole, while the right one
points towards the South Galactic Pole.} \label{fig:2df3d}
\end{figure}

\begin{figure}
\centering
\resizebox{\textwidth}{!}{\includegraphics*{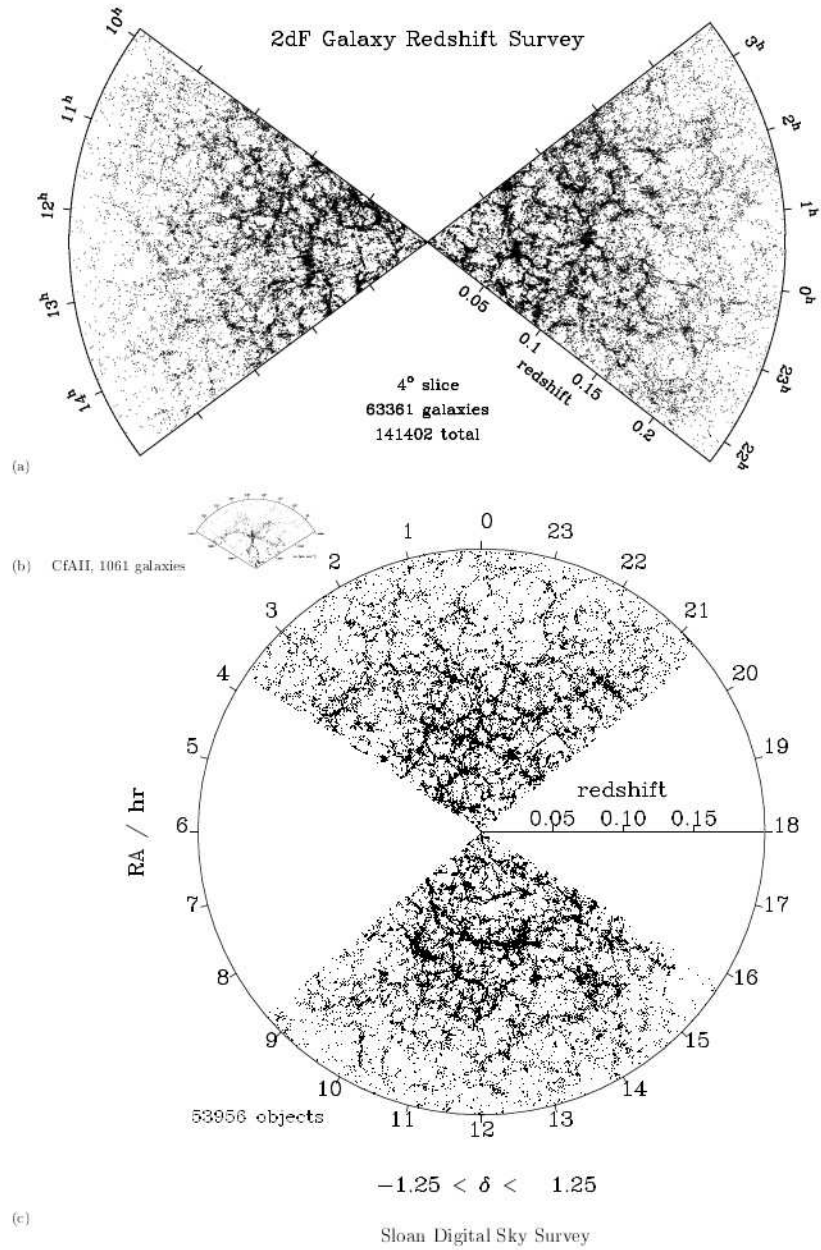}}
\caption{The top diagram shows two slices of $4^\circ$ width and
depth $z=0.25$ from the 2dF galaxy redshift survey, from
\cite{VMpeacock01}. The circular diagram at the bottom has a radius
corresponding to redshift $z=0.2$ and shows 55,958 galaxies from the
SDSS survey, from \cite{VMloveday}). As an inset, the first CfA
slice from \cite{VMlapparent86} is shown to scale.}
\label{fig:slices}
\end{figure}

The present wide field surveys are much deeper as it can be
appreciated in Fig.~\ref{fig:2df3d} and in Fig.~\ref{fig:slices}.
Fig.~\ref{fig:2df3d} illustrates our local neighbourhood (up to $400
\hMpc$) from the Two-Degree Field Galaxy Redshift Survey (2dFGRS)
\index{survey!2dFGRS} in a three dimensional view, where large
superclusters surround more empty regions, delineated by long
filaments. Fig.~\ref{fig:slices} shows the first CfA slice
\index{survey!CfA} with cone diagrams from the 2dFGRS and the Sloan
Didital Sky Survey (SDSS). \index{survey!SDSS} The first one
contains redshifts of about 250,000 galaxies in wide regions around
the north and south Galactic poles with a median redshift $z=0.11$.
It extends up to $z\simeq3$. Galaxies in this survey go down to
apparent blue magnitude $b_{\lim}=19.45$, therefore this is a
magnitude-limited survey that misses faint galaxies at large
distances, as it can be seen in  Fig.~ \ref{fig:slices}. The SDSS
survey is also magnitude-limited, but the limit has been selected to
be red, $r_{\lim}=17.77$. The present release of the SDSS (DR6)
covers an area almost five times as big as the area covered by the
2dFGRS.

More information about these surveys can be found in their web
pages: \url{http://www.mso.anu.edu.au/2dFGRS/} for the 2dF survey
and \url{http://www.sdss.org/} for the SDSS survey.

\section{The Two-point Correlation Function \label{sec:vmcorrfunc}}

\begin{figure}
\centering
\resizebox{0.4\textwidth}{!}{\includegraphics*{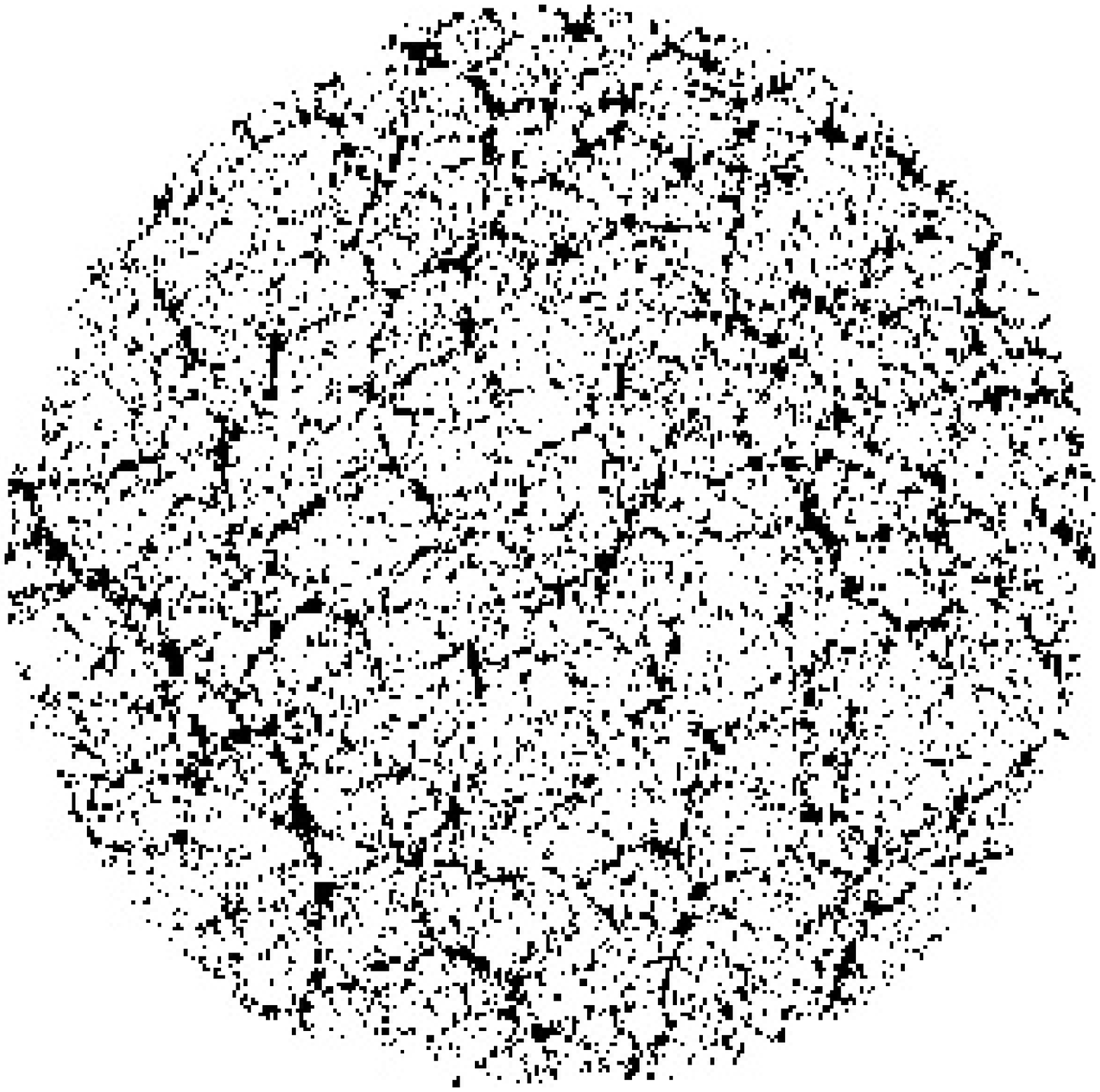}}
\resizebox{0.4\textwidth}{!}{\includegraphics*{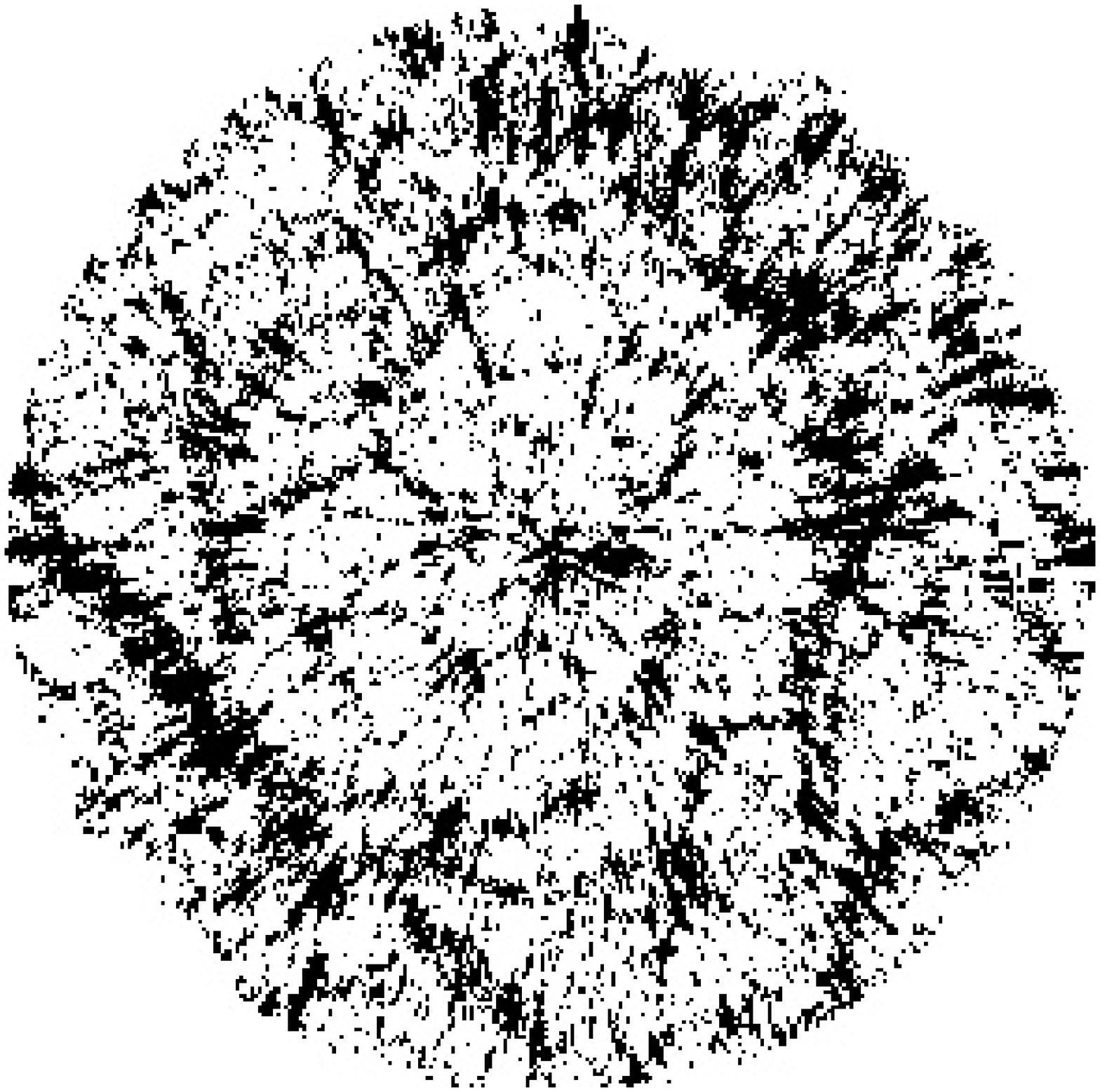}}
\caption{Illustration by a two-dimensional simulation of the effect
of the peculiar velocities distorting the real space (left panel) to
produce the redshift space (right panel). Figures courtesy of Adrian
Melott.} \label{fig:melott}
\end{figure}

After measuring the two-point correlation function
\index{correlation!function!two-point} over projected galaxy
samples, the great challenge was to do it directly for redshift
surveys where the distance inferred from the recession velocities
was used, providing a three-dimensional space. As it has been
already mentioned, we have to bear in mind that measured redshifts
are contaminated by the peculiar velocities. This 3D space, the
so-called redshift space, is a distorted view of the real space.
Fig. ~\ref{fig:melott} shows a simulation with the effect of the
peculiar velocities distorting the real space (left panel),
squeezing the structures to produce the radial stretched structures
pointing to the observer, known as fingers of God (right panel). For
the details see the web page
\url{http://kusmos.phsx.ku.edu/~melott/redshift-distortions.html}.
These fingers of God appear strongest where the galaxy density is
largest, and are attributable to the extra ``peculiar" (ie:,
non-Hubble) component of the velocity of individual galaxies in the
galaxy clusters \cite{VMjackson72, VMsargent77, VMkaiser87,
VMhamilton98} .

Considering two infinitesimal volume elements $dV_1$ and $dV_2$
separated by a vector distance $\mathbf{r}_{12}$, the joint
probability of there being a galaxy lying in each of these volumes
is:
\begin{equation}
dP_{12} = n^2 [1 + \xi(\mathbf{r}_{12})] dV_1 dV_2,
\label{eq:xidef1}
\end{equation}

Assuming homogeneity (the point process is invariant under
translation) and isotropy (the point process is invariant under
rotation) for the galaxy distribution, the quantity depends only on
the distance $r_{12}=|\mathbf{r}_{12}|$ and Eq.~(\ref{eq:xidef1})
becomes Eq.~(\ref{eq:xidef2}).

Apart of the formal definitions given in the previous equations, to
estimate the correlation function for a particular complete galaxy
sample with $N$ objects, several formulae providing appropriate
estimators have been introduced. The most widely used are the
Hamilton estimator \cite{VMhamilton93},
\index{correlation!function!estimator} and the Landy and Szalay
estimator \cite{VMlandy93}. For both, a Poisson catalog, a binomial
process with $N_{\rm{rd}}$ points, has to be generated within the
same boundaries of the real data set. The estimators can be written
as:

\begin{equation}
\widehat{\xi}_{\rm{HAM}} (r) = {\frac{DD(r)\cdot RR(r)} {[DR(r)]^2}} - 1,
\label{eham}
\end{equation}
\begin{equation}
\widehat{\xi}_{\rm{LS}} (r) = 1 +
    \left(\frac{N_{\rm{rd}}}{N}\right)^2
    \frac{DD(r)}{RR(r)}
    - 2 \frac{N_{\rm{rd}}}{N} \frac{DR(r)}{RR(r)}
\label{els}.
\end{equation}
where $DD(r)$ is the number of pairs of galaxies of the data sample
with separation within the interval $[r-dr/2, r+dr/2]$, $DR(r)$ is
the number of pairs between a galaxy and a point of the Poisson
catalog, and $RR(r)$ is the number of pairs between points from the
Poisson catalog \cite{VMpons99,VMkerscher00}.

As it has been explained in the contributions by Hamilton and
Szapudi in this volume, the above formulae have to be corrected due
to the selection effects. These effects could be radial due to the
fact that redshift surveys are built as apparent magnitude catalogs,
and therefore fainter galaxies are lost at larger distances, and
could be angular due to the Galactic absorption that makes the sky
not equally transparent in all directions or to the fact that
different areas of the sky within the sample boundaries are not
equally covered by the observations, therefore providing varying
apparent magnitude limit depending on the direction. Moreover some
areas could not be covered at all because of the presence of nearby
stars, or because of fiber collisions in the spectrograph. In order
to account for this complexity the best solution is to use the
freely available MANGLE software
(\url{http://space.mit.edu/home/tegmark/mangle/}), a generic tool
for managing angular masks on a sphere \cite{VMmangle}.

\subsection{The projected correlation function}

\begin{figure}
\centering
\resizebox{0.6\textwidth}{!}{\includegraphics*{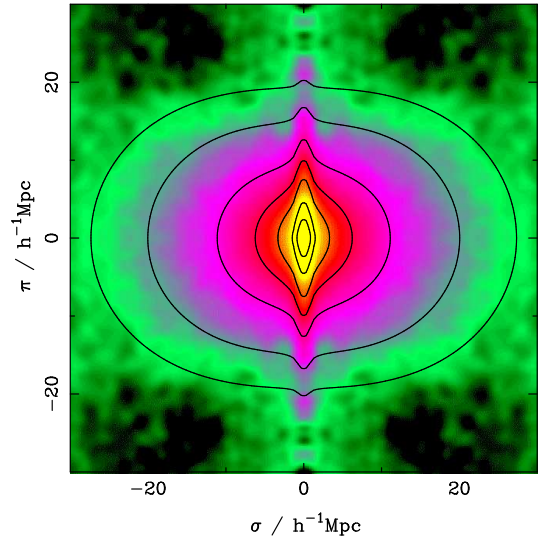}}
\caption{The galaxy correlation function $\xi(\pi, r_p)$ for the
2dFGRS (transverse distance $r_p$ is represented here by $\sigma$).
This diagram show the two sources of anisotropy in the correlation
function: the radial smearing due to random velocities within groups
and clusters at small distances and the large scale flattening
produced by coherent infall velocities. In this diagram the
calculation has been performed by counting pairs in boxes and then
smoothing with a Gaussian. The results obtained for the first
quadrant are repeated with reflection in both axes to show
deviations from circular symmetry. Overplotted lines correspond to
the function calculated for a given theoretical model. Figure from
\cite{VMpeacock01}.} \label{fig:xipisig}
\end{figure}

Since at small scales, peculiar velocities strongly distort the
correlation function, it has become customary to calculate and
display the so-called projected correlation function
\index{correlation!function!projected}
\begin{equation}
\label{eq:wp}
w_p (r_p) = 2 \int_0^{\infty} \xi(\pi, r_p) \mathrm{d}\pi,
\end{equation}
where the two-dimensional correlation function $\xi(\pi, r_p)$ is
computed on a grid of pair separations parallel ($\pi$) and
perpendicular ($r_p$) to the line of sight. Fig.~\ref{fig:xipisig}
shows this function calculated by Peacock et al. \cite{VMpeacock01}
for the 2dFGRS.

\begin{figure}
\centering
\resizebox{0.4\textwidth}{!}{\includegraphics*{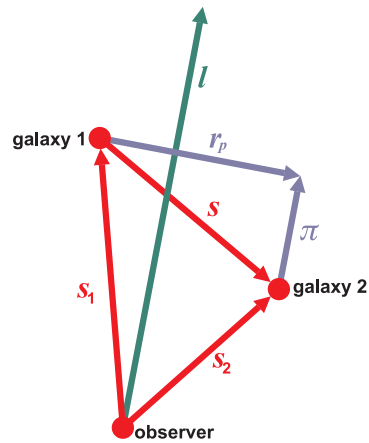}}
\caption{Illustration of the parallel and perpendicular separations
between two objects.} \label{fig:ilus}
\end{figure}

If the separation vector between two positions in redshift space is
$\vec{s} = \vec{s_2} - \vec{s_1}$, and the line-of-sight vector is
$\vec{l} = \vec{s_1} + \vec{s_2}$, the parallel and perpendicular
distances of the pair are (see Fig.~\ref{fig:ilus}):
\[
\pi = \frac{|\vec{s} \cdot \vec{l}|}{|\vec{l}|} ,
\quad r_p = \sqrt{\vec{s} \cdot \vec{s} - \pi ^2 }.
\]

\begin{figure}
\centering
\resizebox{\textwidth}{!}{\includegraphics*{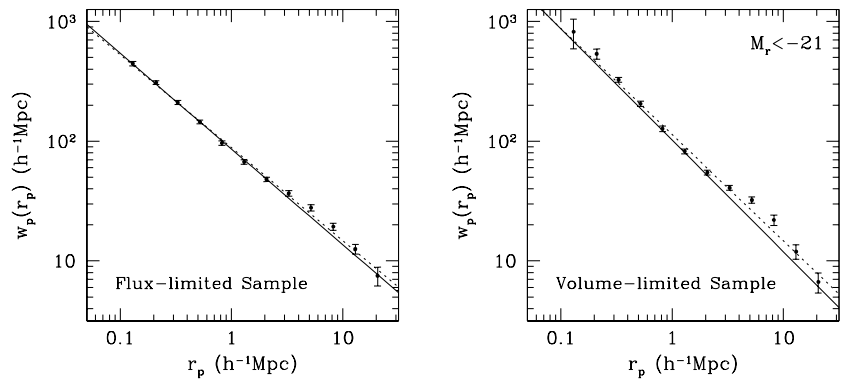}}
\caption{The projected correlation function $w_p(r_p)$ for the SDSS
data. Left panel shows the result for the flux limited sample and
right panel for the volume-limited sample. Two different power-law
fits to the data have been performed. Solid lines make use of the
full covariance matrix while dashed lines only use the diagonal
elements. Figure from Zehavi et al. \cite{VMzehavi04}.}
\label{fig:wp}
\end{figure}

Fig.~\ref{fig:wp} shows the projected correlation function
calculated for the Sloan Digital Sky Survey by Zehavi et al.
\cite{VMzehavi04}. The relation between the projected correlation
function and the three-dimensional real correlation function (not
affected by redshift distortions) is, for an isotropic distribution
\cite{VMDP83}:
\begin{equation}
\label{eq:isotropy}
w_p (r_p) = 2 \int_{r_p}^{\infty} \xi(r)
\frac{r \mathrm{d}r}{\left( r^2 - r_p^2 \right) ^{1/2}}.
\end{equation}
From the previous equation it is straightforward to see that if
 $\xi(r)$ fits well a power law, i.e. $\xi(r) = (r / r_0 )^{-\gamma}$,
$w_p (r_p)$ also does, $w_p(r_p) = A r_p^{-\alpha}$, with
\[
\alpha = \gamma - 1, \quad \mbox{and} \quad A = \frac{r_0^{\gamma} \Gamma(0.5)
\Gamma[0.5(\gamma -1)] }{ \Gamma(0.5\gamma)}  .
\]
In practice, the integration in Eq.~\ref{eq:wp} is performed up to a
fixed value $\pi_{\max}$ which depends on the survey. For the SDSS,
Zehavi et al. \cite{VMzehavi04} used $\pi_{\max} = 40 \hMpc$, a
value considered large enough by the authors to include the relevant
information to measure $w_p(r_p)$ in the range $0.1 \hMpc < r_p < 20
\hMpc$. The assumed cosmological model for the calculation of
distances is the concordance model for which $\Omega_m=0.3$ and
$\Omega_\Lambda=0.7$.

The function shown in the left panel of Fig.~\ref{fig:wp} has been
calculated making use of a subset containing 118,149 galaxies drawn
from the flux-limited sample selected by Blanton et al.
\cite{VMblanton03}. The estimator of the correlation function makes
use of the radial selection function that incorporates the
luminosity evolution model of Blanton et al. \cite{VMblanton03}. On
the right panel the calculation has been performed over a
volume-limited sample containing only galaxies bright enough to be
seen within the whole volume (up to $ 462 \hMpc$, the limit of the
sample). This subsample contains 21,659 galaxies with absolute red
magnitude $M_r < -21$ (for $h=1$). The solid line on the left panel
of Fig.~\ref{fig:wp} shows the fit to $w_p(r_p)$ which corresponds
to a real-space correlation function $\xi(r) = (r / 5.77
\hMpc)^{-1.80}$. For the volume-limited sample the fit shows a
slightly steeper power-law $\xi(r) = (r / 5.91 \hMpc)^{-1.93}$.
\index{correlation!function!power-law} This is a expected
consequence of the segregation of luminosity as we will show later,
since galaxies in this subsample are 0.56 magnitudes brighter than
the characteristic value of the Schechter \cite{VMschechter76}
luminosity function \cite{VMblanton03}.

Although it is remarkable from the power-law fits shown in
Fig.~\ref{fig:wp} how the scaling holds for about three orders of
magnitude in scale, the main point stressed in this analysis was
precisely the unambiguous detection of a systematic departure from
the simple power-law behaviour. A similar result  was also obtained
by Hawkins et al. \cite{VMhawkins03} for the 2dfGRS, although the
best fit power-law for the correlation function of 2dF galaxies is
$\xi(r) = (r / 5.05 \hMpc)^{-1.67}$ with a less steep slope than the
one found for SDSS galaxies and with a value of the correlation
length $r_0=5.05 \pm 0.26 \hMpc$, substantially smaller than the
SDSS result. Again, this can be explained as a consequence of the
different galaxy content of both surveys, SDSS are red-magnitude
selected while 2dF are blue-magnitude selected.

Error bars for the correlation function in Fig.~\ref{fig:wp} have
been calculated in two different ways which illustrate the two main
methods currently used. For the flux-limited sample, jackknife
resampling \index{resampling!jackknife} of the data has been used.
The sample is divided into $N$ disjoint subsamples covering each
approximately the same area of the sky, then the calculation of
$\xi(r)$ is performed on each of the jackknife samples created by
summing up the $N$ subsamples except one, which is omitted in turn.
The $ij$ element of the covariance matrix is computed by
\cite{VMzehavi02}
\begin{equation}
\label{eq:jk}
C_{ij} = \frac{N-1}{N} \sum_{k=1}^{N}
({\xi_i}^k - {\bar{\xi}_i})({\xi_j}^k - {\bar{\xi}_j}),
\end{equation}
where $\bar{\xi}_i$ is the average value of  $\xi_i$ measured on the
jackknife samples. Statistical errors can be calculated using the
whole covariance matrix, or just making use of the elements in the
diagonal, and thus ignoring the correlation amongst the errors. The
other possibility consists in using mock catalogues from N-body
simulations or semi-analytical models of structure formation with a
recipe for allocating galaxies. These mock catalogues can be used as
the subsamples in which Eq.~\ref{eq:jk} can be applied to obtain the
covariance matrix.

The variation of the slope in the two-point correlation function of
galaxies with the scale might be ascribed to the existence of two
different clustering regimes: the small scale regime dominated by
pairs of galaxies within the same dark matter halo and a second
regime where pairs of galaxies belonging to different halos
contribute to the downturn of the power-law in $w_p(r_p)$.

\subsection{Galaxy properties and clustering}

The photometric and spectral information provided by surveys like
SDSS and 2dFGRS allows to study how the clustering of galaxies
depends on different factors such as luminosity, morphology, colour
and spectral type, although these factors are certainly not
independent. For example, it is well known
\cite{VMdavis76,VMdressler80} that early-type galaxies show more
pronounced clustering at small separations than late-type galaxies,
the first kind displaying steeper power-law fits to their
correlation than the latter. This segregation plays an interesting
role in the understanding of the galaxy formation process, since
galaxies are biased tracers of the total matter distribution in the
universe (mainly dark) and the bias also depends on the scale
\cite{VMlahav02}. Madgwick et al. \cite{VMMadgwick03} have recently
divided the 2dFGRS in two subsets: passive galaxies with a
relatively low star formation rate, and active galaxies with higher
current star formation rate. This division correlates well with
colour and morphology, being passive galaxies mainly red old
ellipticals.

\begin{figure}
\centering
\resizebox{0.42768\textwidth}{!}{\includegraphics*{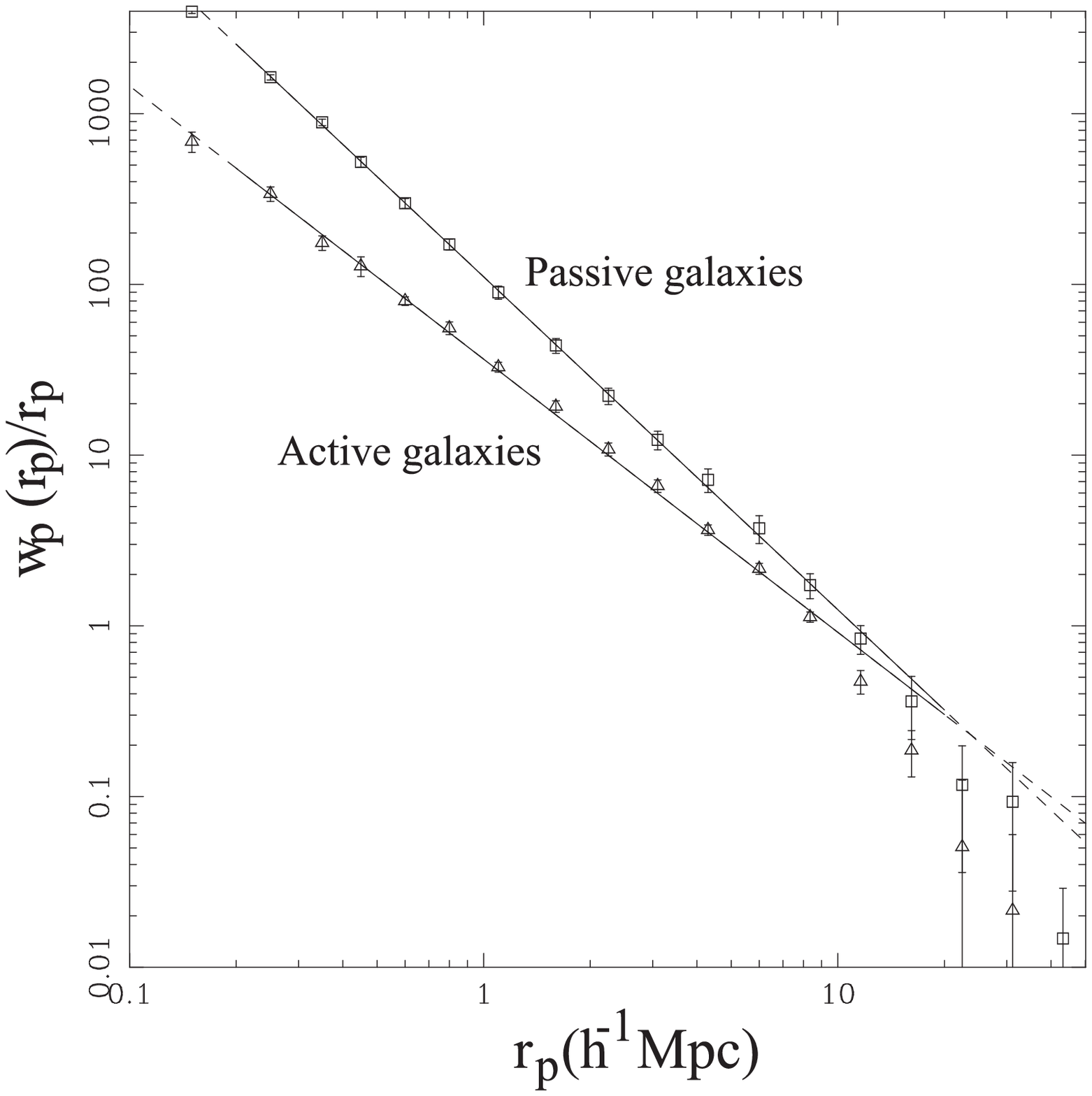}}
\resizebox{0.55836\textwidth}{!}{\includegraphics*{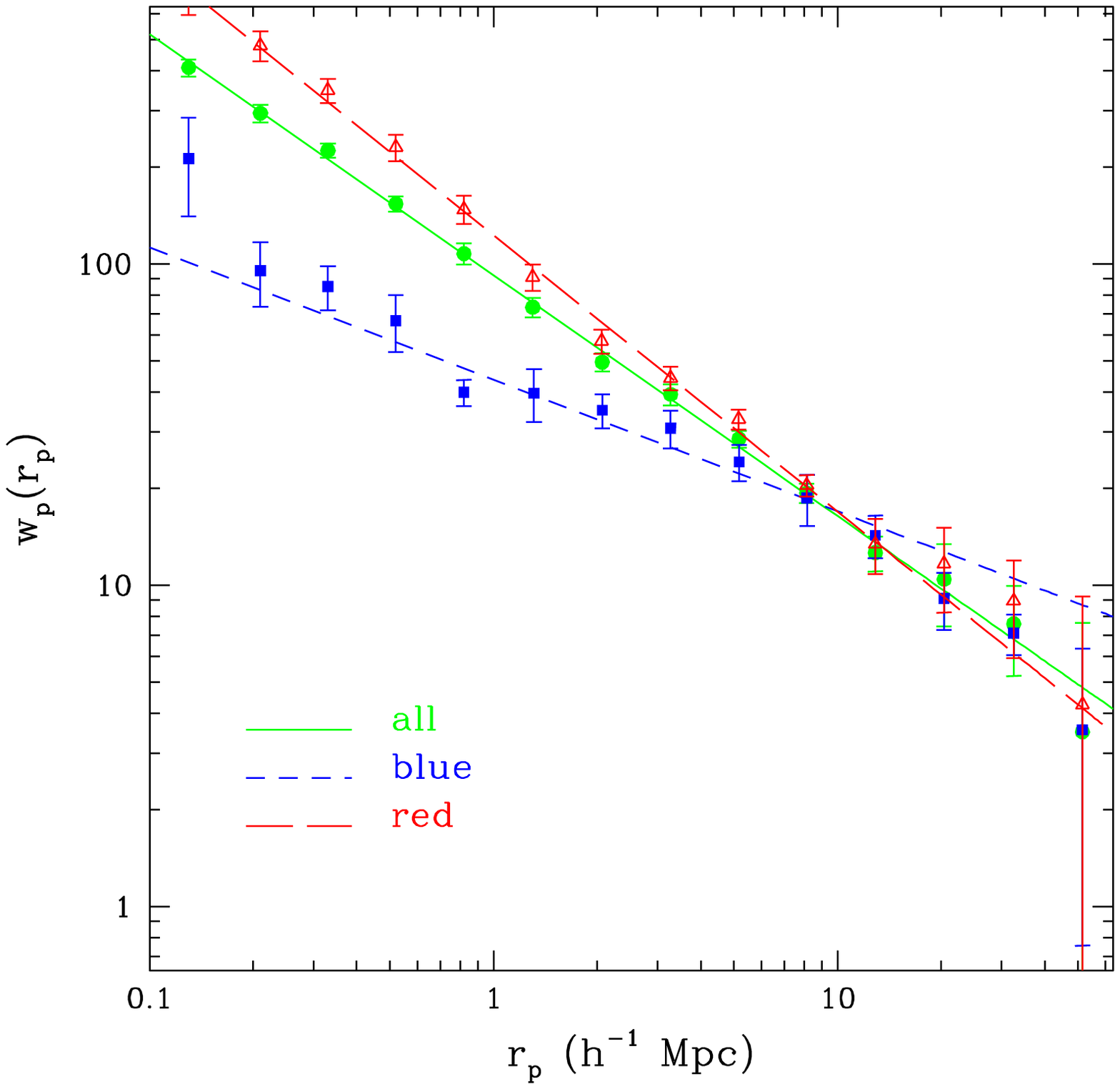}}
\caption{In the left panel, we show the projected correlation
function $w_p(r_p)$ for two subsamples of the 2dfGRS data where the
division has been performed in terms of current star formation rate.
Passive galaxies cluster stronger than their active counterparts.
Figure adapted from Madgwick et al. \cite{VMMadgwick03}. In the
right panel, it is shown the projected correlation function of
subsamples divided by colour drawn from the SDSS. Different lines
show the best-fit power-laws for $w_p(r_p)$. The short-dashed,
long-dashed and solid lines correspond to the blue, red, and full
samples, respectively. Figure from Zehavi et al. \cite{VMzehavi02}.}
\label{fig:segre}
\end{figure}

Fig.~\ref{fig:segre} (left panel) shows the projected
correlation function for these two subsets. As it can be
appreciated, passive galaxies present a two-point correlation
function with steeper slope and larger amplitude than active
galaxies, being the best fit for each subsample $\xi(r) = (r / 6.10
\pm 0.34 \hMpc)^{-1.95\pm0.03}$ for passive galaxies and $\xi(r) =
(r / 3.67\pm0.30 \hMpc)^{-1.60\pm0.04}$ for active galaxies. A
similar analysis was also performed by Zehavi et al.
\cite{VMzehavi02} dividing an early release of the SDSS galaxies
into two subgroups by colour, red and blue, using the value of the
colour $u^*-r^*=1.8$ for the division.  The blue subset contains
mainly late morphological types while the red group is formed mainly
by bulge galaxies, as it should be expected. Again, as it can be
appreciated in Fig.~\ref{fig:segre} (right panel), red galaxies
cluster stronger than blue galaxies, being their best fit to a
power-law \index{correlation!function!power-law} in the range $[0.1
\hMpc < r_p < 16 \hMpc]$, $\xi(r) = (r / 6.78 \pm 0.23
\hMpc)^{-1.86\pm0.03}$, while for blue galaxies the best fit is
$\xi(r) = (r / 4.02\pm0.25 \hMpc)^{-1.41\pm0.04}$. Blanton et al.
\cite{VMblanton05} have shown that large amplitudes in the
correlation function corresponding to subsets selected by luminosity
or colour are typically accompanied with steeper slopes.

\section{The Power Spectrum \label{sec:vmpower}}

The power spectrum $P(k)$ is a clustering descriptor depending on
the wavenumber $k$ that measures the amount of clustering at
different scales. It is the Fourier transform of the correlation
function, and therefore both functions contain equivalent
information, although it can be said that they describe different
sides of the same process. For a Gaussian random field, the Fourier
modes are independent, and the field gets completely characterised
by its power spectrum. As the initial fluctuations from the
inflationary epoch in the universe are described as a Gaussian
field, the model predictions in Cosmology are typically made in
terms of power spectra.

\index{power spectrum} The Power spectrum and the correlation
function are related through a Fourier transform:
\[
P(\mk)=\int \xi(\mr)e^{i\mk\cdot\mr}\,d^3r,
\]
\[
\xi(\mr)=\int P(\mk)e^{-i\mk\cdot\mr}\,\frac{d^3k}{(2\pi)^3},
\]
Assuming isotropy, the last equation can be rewritten as:
\[
\xi(r)=4\pi\int_0^\infty P(k)\frac{\sin(kr)}{kr}\frac{k^2\,dk}{(2\pi)^3}.
\]
Some authors \cite{VMpeacock99} prefer to use the following
normalization for the power spectrum:
\[
\Delta^2(k)=\frac{1}{2\pi^2}P(k)k^3,
\]
in such a way that the
total variance of the density field is just:
\[
\sigma_\delta^2=\int_0^\infty\Delta^2(k)\,d(\ln k).
\]

One of the advantages of the power spectrum over the correlation
function is that amplitudes for different wavenumbers are
statistically orthogonal (for a more detailed discussion see the
contributions by Andrew Hamilton in this volume):
\begin{equation}
E\left\{\widetilde{\delta}(\mk)\widetilde{\delta}^\star(\mk')\right\}=
    (2\pi)^3\delta_D(\mk-\mk')P(\mk).
\end{equation}
Here $\widetilde{\delta}(\mk)$ is the Fourier amplitude of the
overdensity field $\delta=(\rho-\bar{\rho})/\bar{\rho}$ at a
wavenumber $\mk$, $\rho$ is the matter density, a star denotes
complex conjugation, $E\{\}$ denotes expectation values over
realizations of the random field, and $\delta_D(\mx)$ is the
three-dimensional Dirac delta function.

If we have a sample (catalog) of galaxies with the coordinates
$\mx_j$, we can write the estimator for a Fourier amplitude of the
overdensity distribution \cite{VMfkp} (for a finite set of
frequencies $\mk_i$) as
\[
F(\mk_i)=\sum_j\frac{\psi(\mx_j)}
    {\bar{n}({\mx}_j)}e^{i\mk_i\cdot\mx} -\widetilde{\psi}(\mk_i),
\]
where $\bar{n}(\mx)$ is the position-dependent selection
function (the observed mean number density) of the sample and
$\psi(\mx)$ is a weight function that can be selected
at will.

The raw estimator for the spectrum is
\[
P_R(\mk_i)=F(\mk_i)F^\star(\mk_i),
\]
and its expectation value
\[
E\left\{\langle|F(\mk_i)|^2\rangle\right\}
    =\int G(\mk_i-\mk')P(\mk')\,\frac{d^3k'}{(2\pi)^3}
    +\int_V\frac{\psi^2(\mx)}{\bar{n}(\mx)}\,d^3x,
\]
where $G(\mk)=|\tilde{\psi}(\mk)|^2$ is the window function that
also depends on the geometry of the sample volume. The reader can
learn more about the estimation of the power spectrum in the
contributions by Andrew Hamilton in this volume.

\subsection{Acoustic peak in $\xi$ and acoustic oscillations in $P(k)$}

\index{acoustic!peak}\index{acoustic!oscillations}

Prior to the epoch of the recombination, the universe is filled by a
plasma where photons and baryons are coupled. Due to the pressure of
photons, sound speed is relativistic at this time and the sound
horizon has a comoving radius of 150 Mpc. Cosmological fluctuations
produce sound waves in this plasma.

\begin{figure}
\centering
\resizebox{\textwidth}{!}{\includegraphics*{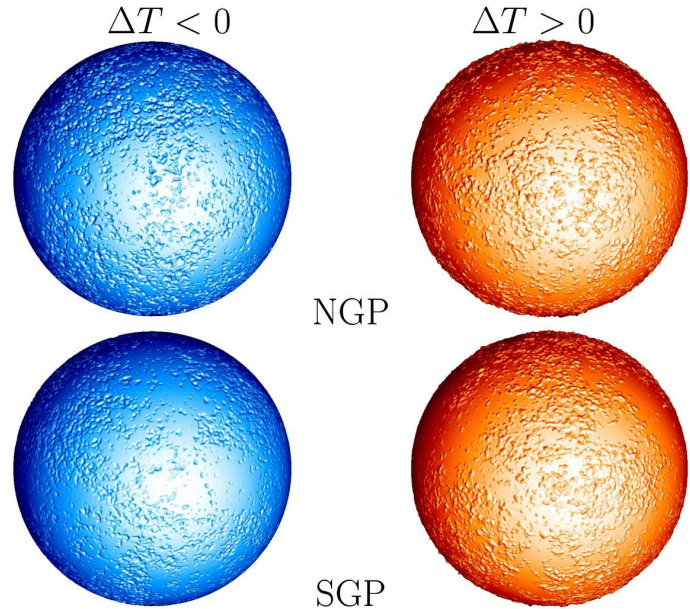}}
\caption{Temperature fluctuations of the WMAP data. The two upper
spheres are centred in the north Galactic pole (NGP), while the
bottom two are in the south Galactic pole (SGP). On the left hand
side, in blue, pixels where $\Delta T < 0$ are depicted as depths,
while on the right hand side, red pixels with  $\Delta T > 0$ are
displayed as elevations. The ``sea level" in blue spheres
corresponds to the pixels where  $\Delta T \ge 0$ and in the red
sphere, where  $\Delta T \le 0$.} \label{fig:balls}
\end{figure}

At about 380,000 years after the Big Bang, when the temperature has
fallen down to 3000 K, and recombination takes place, the universe
loses its ionized state and neutral gas dominates. At this state,
sound speed drops off abruptly and acoustic oscillations in the
fluid become frozen. Their signature can be detected in both the
Cosmic Microwave Background (CMB) radiation and the large-scale
distribution of galaxies. Fig.~\ref{fig:balls} shows a
representation of the temperature at the last scattering surface
from WMAP. These fluctuations have been analyzed in detail to obtain
a precise estimation of the anisotropy power spectrum of the CMB.
The acoustic peaks in this observed angular power spectrum (see
contribution by Enrique Mart\'{\i}nez-G\'onzalez in this volume)
have become a powerful cosmological probe. In particular, the CMB
provides an accurate way to measure the characteristic length scale
of the acoustic oscillations, that depends on the speed of sound,
$c_s$, in the photon-baryon fluid and the cosmic time when this
takes place. The distance that a sound wave has traveled at the age
of the universe at that time is
\begin{equation}
\label{eq:soundh}
 \int_0^{t_{rec}} c_s(t) (1+z) dt \simeq 147 \pm 2 \,\, \mbox{Mpc}.
\end{equation}
for the standard flat $\Lambda$-CDM model. This fixed scale
imprinted in the matter distribution at recombination can be used as
a ``standard ruler" for cosmological purposes.

\begin{figure}
\centering
\resizebox{0.9\textwidth}{!}{\includegraphics*{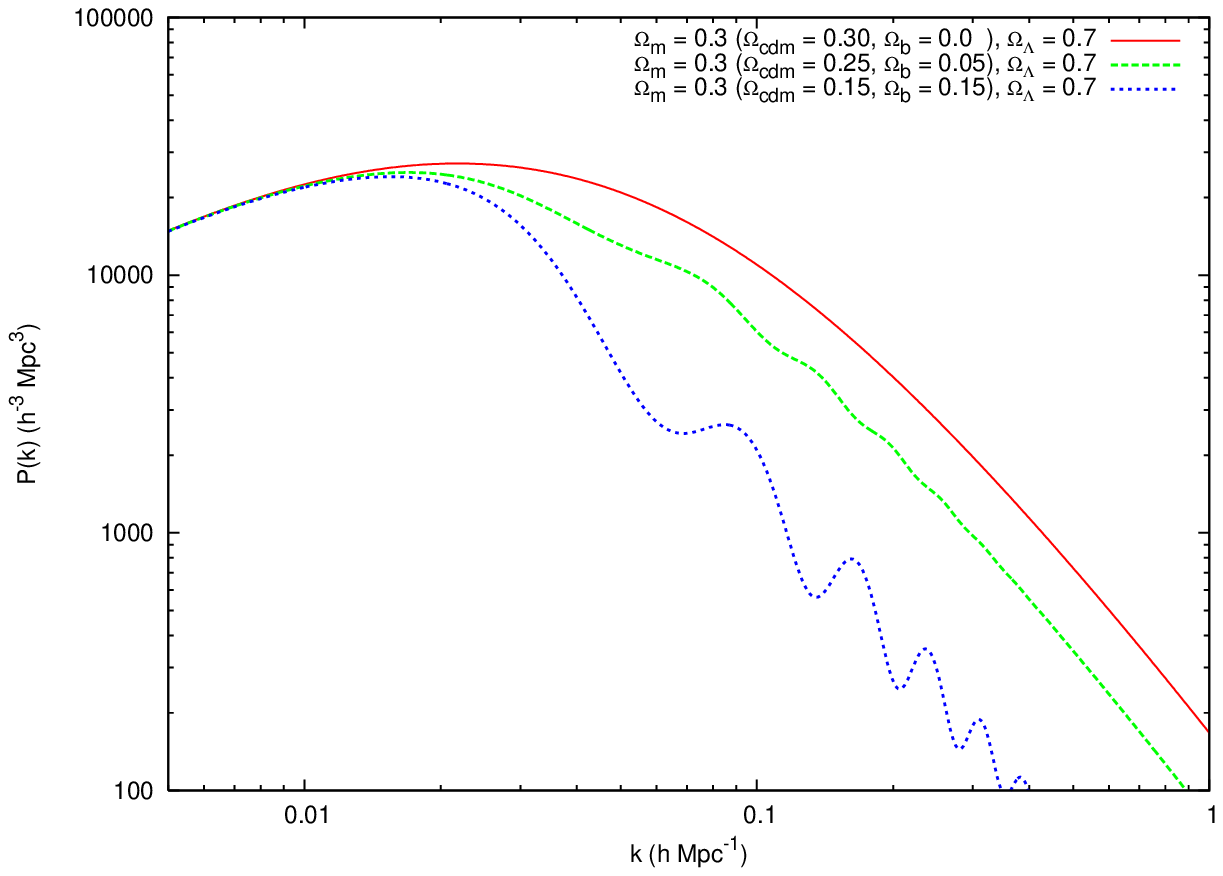}}
\resizebox{0.9\textwidth}{!}{\includegraphics*{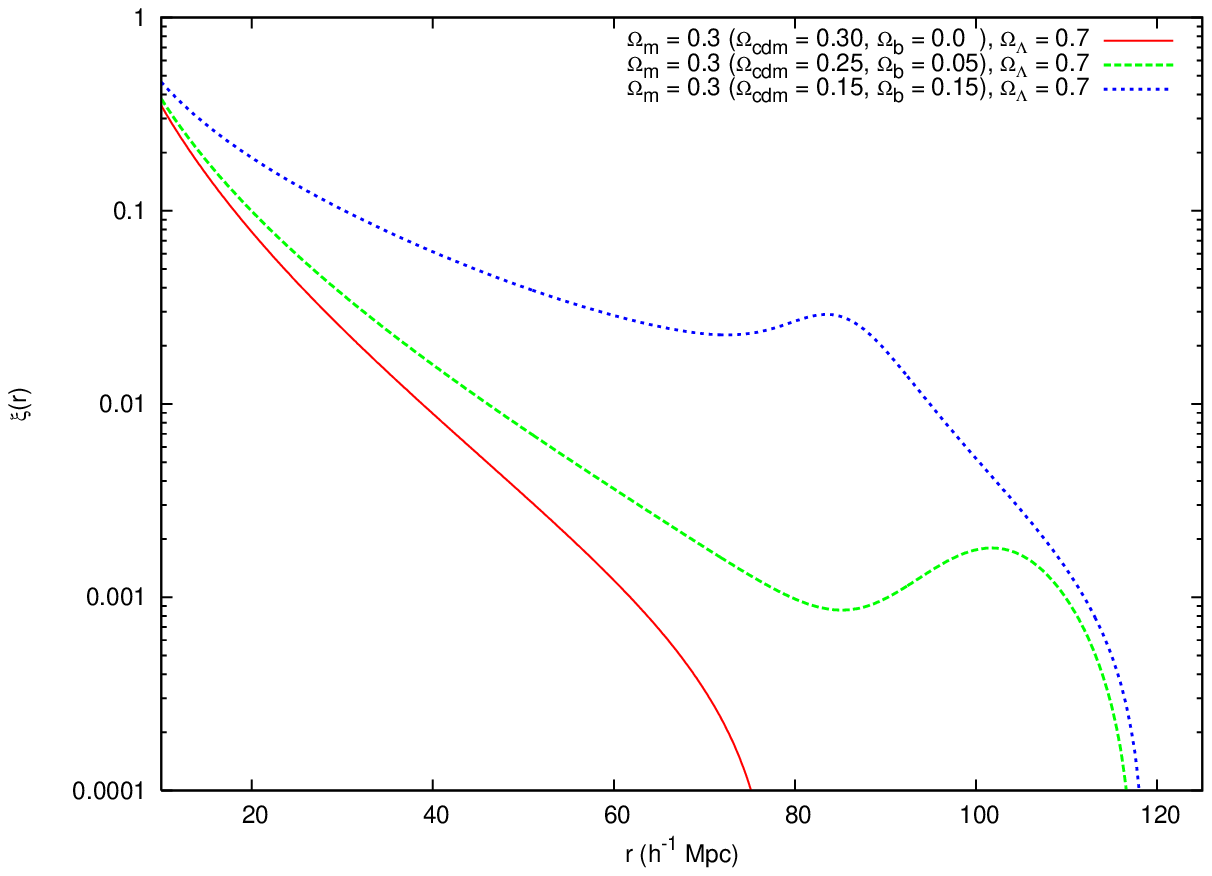}}
\caption{Top panel. The linear-regime power spectrum of the matter
in the universe for different flat models with $\Omega_{\rm
total}=1, \, h=0.7$, \, $\Omega_m=0.3$ and $\Omega_\Lambda=0.7$. The
three curves correspond to different proportions of baryonic and
cold dark matter: from top to bottom $\Omega_b=0, 0.05, 0.15$. As we
see, increasing the baryon content (at fixed $\Omega_m$) increases
the amplitude of the acoustic oscillations, while suppresses power
on small scales (large wavenumber). In the bottom panel, it is shown
the corresponding correlation function to each model displayed with
the same line style. For no baryons (pure cold dark matter), the
acoustic peak is missing, while the peak amplitude is larger with a
larger proportion of baryons. Data for the figure courtesy of Gert
H\"utsi. A similar diagram can be found in \cite{VMnichol07} and
\cite{VMeis98}.} \label{fig:peakxi}
\end{figure}

The imprint in the matter distribution of this acoustic feature
should be detected in both the correlation function and the matter
power spectrum. However, the amplitude of the acoustic peaks in the
CMB angular power spectrum is much larger than the expected
amplitude of the oscillations in the matter power spectrum, which
are called for obvious reasons baryonic acoustic oscillations (BAOs)
\index{acoustic!oscillations!baryonic} Moreover, the feature should
be manifested as a single peak in the correlation function at about
100 $h^{-1}$ Mpc, while in the power spectrum it should be detected
as a series of small-amplitude oscillations as it is shown in
Fig.~\ref{fig:peakxi}. Baryons represent only a small fraction of
the matter in the universe, and therefore, as it can be appreciated
in the figure, the amplitude of the oscillations in the power
spectrum are rather tiny for the concordance model (green dashed
line in the top panel of Fig.~\ref{fig:peakxi}). We can see how
increasing the baryon fraction increases the amplitude of the
oscillations, while wiggles disappear for a pure $\Lambda$-CDM model
(with no baryonic content). At small scales the oscillations are
erased by Silk damping, therefore one needs to accurately measure
the power-spectrum or the correlation function on scales between
$50-150 \, h^{-1}$ Mpc to detect theses features.

\begin{figure}
\centering
\resizebox{0.44\textwidth}{!}{\includegraphics*{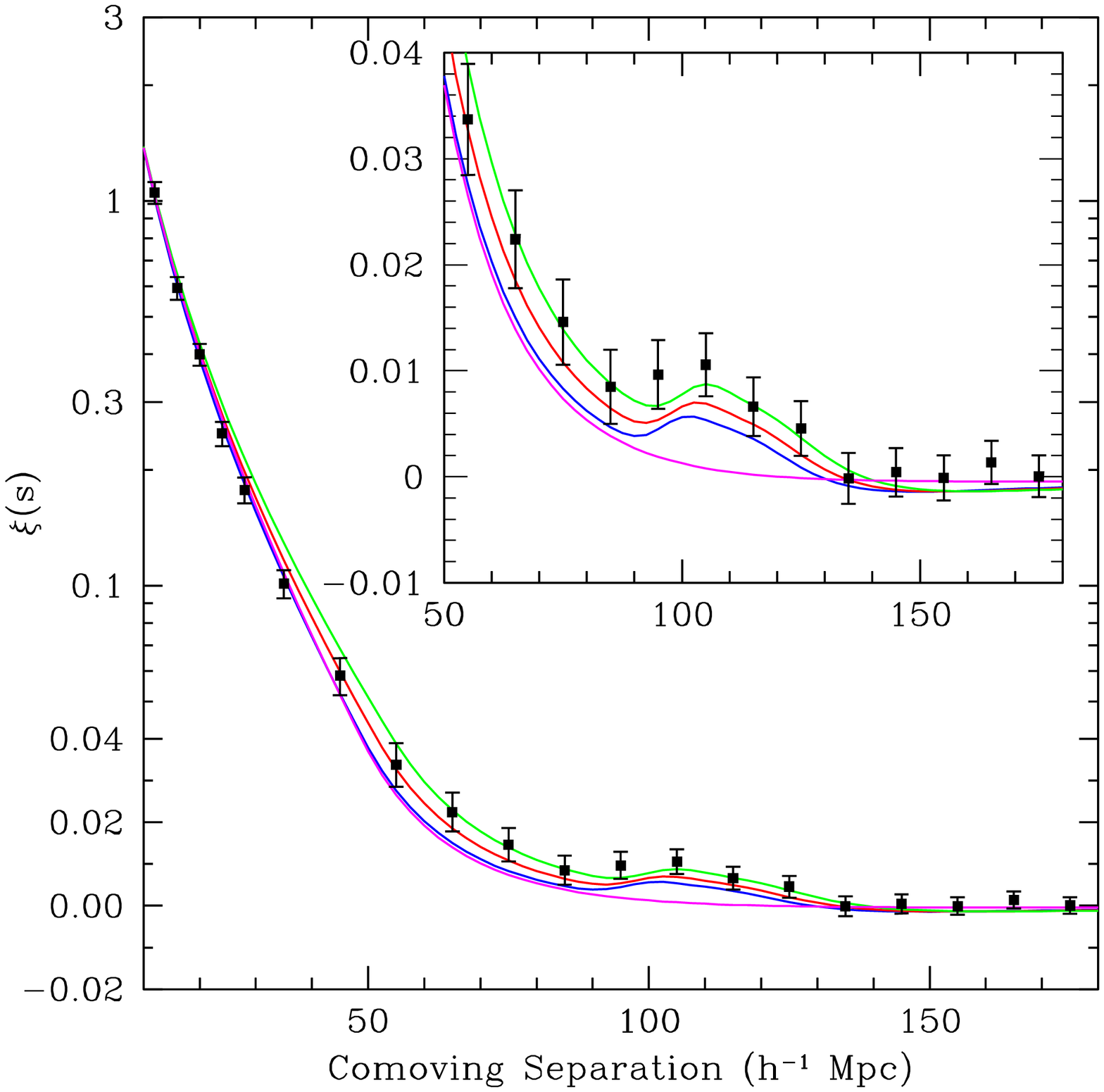}}
\resizebox{0.50\textwidth}{!}{\includegraphics*{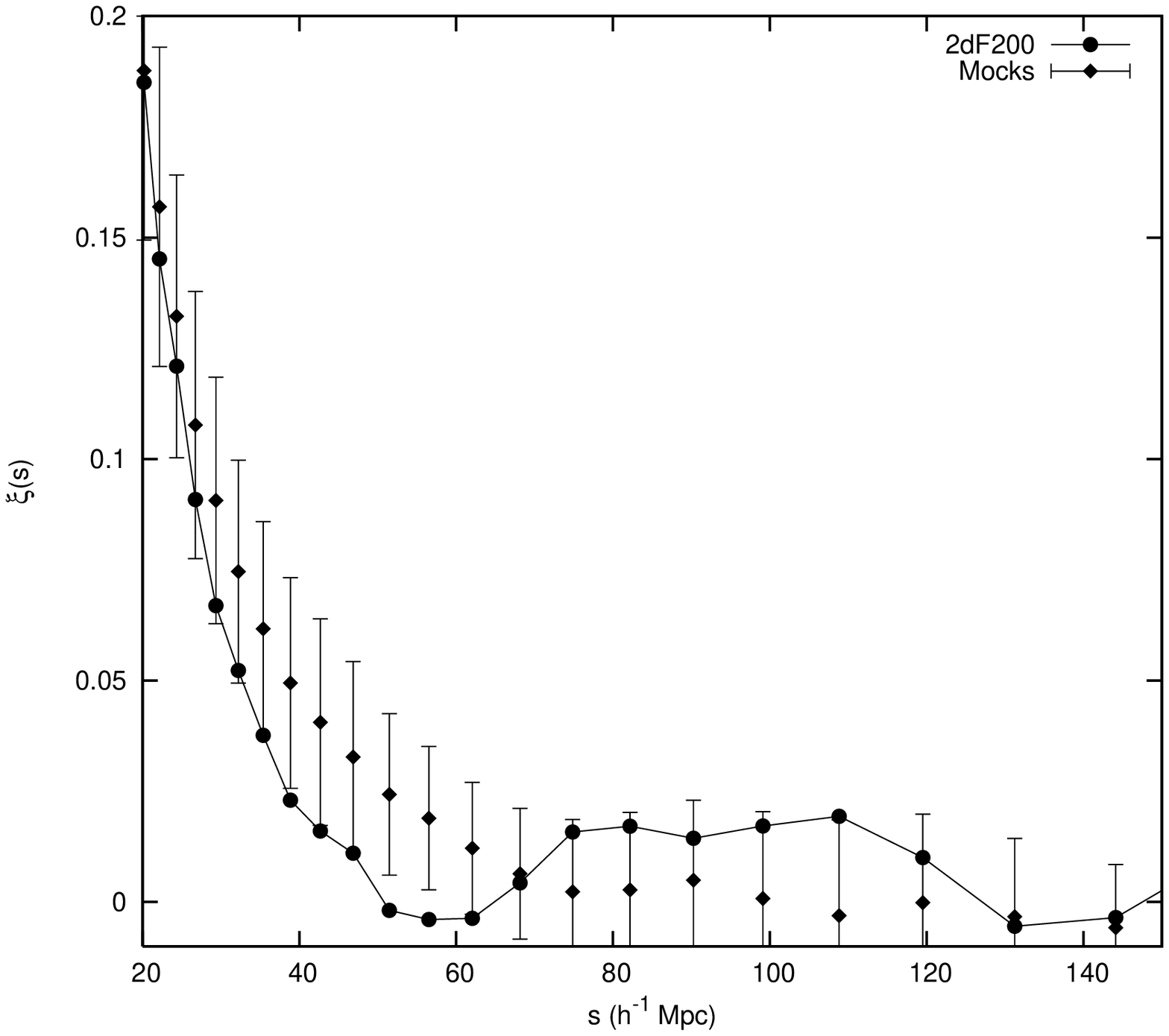}}

\caption{Left. The redshift-space galaxy correlation function
measured for the LRG SDSS sample containing 46,748 luminous read
galaxies in redshift space. The vertical axis mixes logarithmic and
linear scales.The inset shows an expanded view around the peak
($\sim 100 \, h^{-1}$ Mpc) with the vertical axis in linear scale.
The different solid curves correspond to a $\Lambda$-CDM model with
$\Omega_m h^2=0.12$ (green), 0.13 (red), 0.14 (blue); in all cases
the baryon content is fixed to $\Omega_b h^2=0.024$). The magenta
line corresponds to a  pure $\Lambda$-CDM model with no baryons.
Figure from Eisenstein et al. \cite{VMeisenstein05}. Right. The
redshift-space galaxy correlation function measured for a
volume-limited sample extracted from the 2dFGRS (solid discs joined
by a solid line). The same function has been calculated on the 22
mocks models explained in the text. The average correlation function
together with 1-$\sigma$ deviations are shown in the diagram. Mocks
do not show the peak detected in the real galaxy survey.}

\label{fig:lrgpeakxi}
\end{figure}

\begin{figure}
\centering
\resizebox{0.65\textwidth}{!}{\includegraphics*{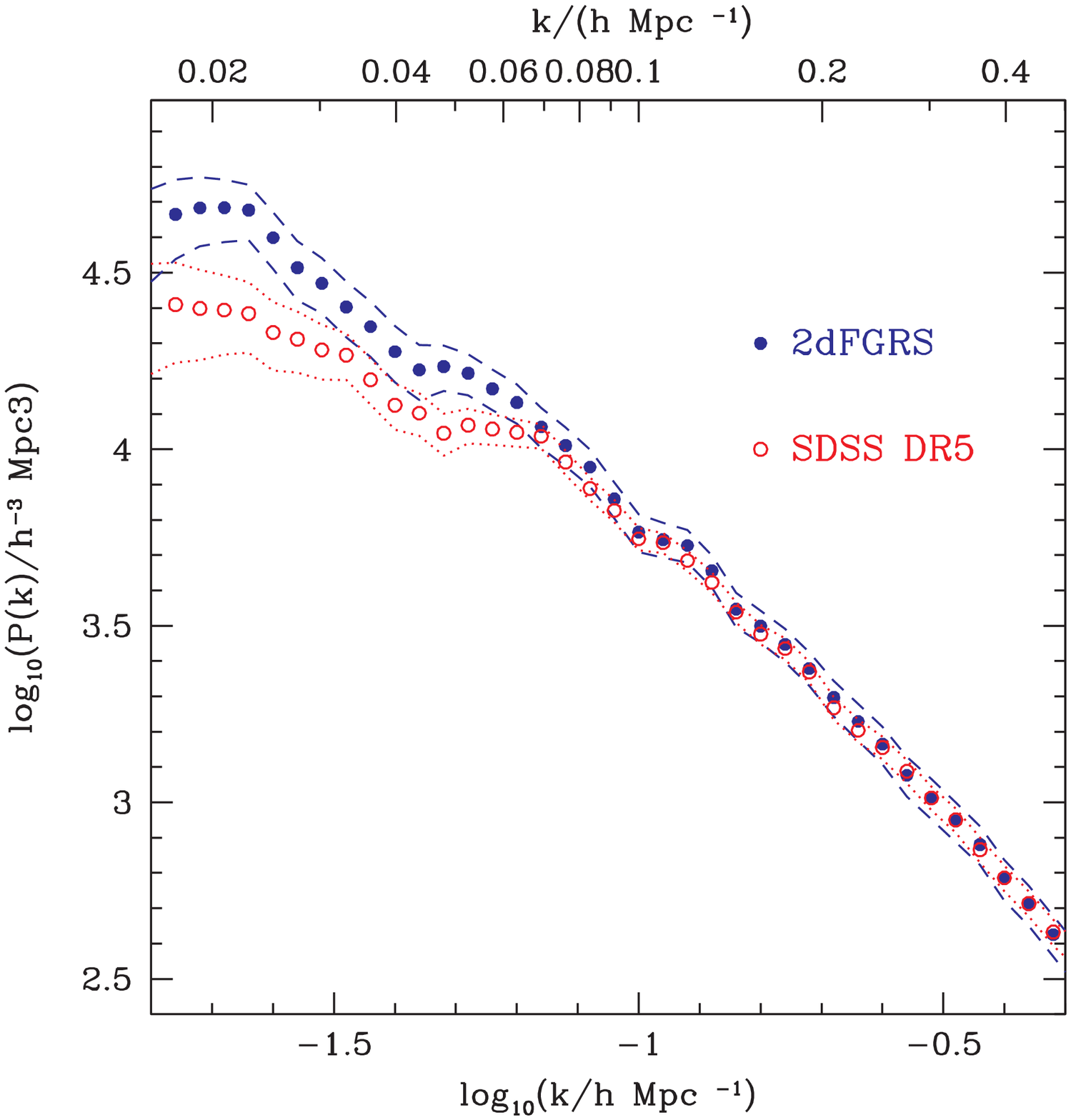}}
\caption{The matter power spectrum $P(k)$ for the 2dFGRS and the
SDSS-DR5. The agreement is good at small scales, while at larger
scales there is a clear evidence of SDSS having more small-scale
power than 2dfGRS.  S\'anchez and Cole \cite{VMsanchez08} interpret
this result as a consequence of the stronger scale-dependent bias
shown by the red galaxies that dominate the SDSS catalogue. Figure
adapted from S\'anchez and Cole \cite{VMsanchez08}.}
\label{fig:pkall}
\end{figure}

Eisenstein et al. (2005) \cite{VMeisenstein05} announced the
detection of the acoustic peak in the two-point redshift-space
correlation function of the SDSS LRG survey (see
Fig.~\ref{fig:lrgpeakxi}). More or less simultaneously, Cole et al.
(2005) \cite{VMcole05} discovered the corresponding feature in form
of wiggles of about 10\% amplitude in the power spectrum of 2dF
galaxy redshift survey. We have also calculated the redshift
correlation function for a nearly volume-limited sample of the
2dFGRS extracted by Croton et al. \cite{VMcroton04}. There are about
25,000 galaxies in this sample with absolute magnitude within the
range $-20 > M_{B_J} - 5\log_{10} h > -21$. The correlation function
displayed in the right panel of  Fig.\ref{fig:lrgpeakxi} shows a
prominent peak around $100 \, h^{-1}$ Mpc which expands for a wider
scale range that the bump observed in the SDSS-LRG sample (left
panel). This could be due to scale-dependent differences between the
clustering of the two samples. A similar effect has been recently
observed in the power spectrum \cite{VMsanchez08} of the two surveys
(see also the figure caption of Fig.~\ref{fig:pkall}). Of course,
the statistical significance of this feature is still to be tested.
Interestingly enough is the fact that the mock catalogues generated
by Norberg et al. \cite{VMnorberg02} to mimic the properties of the
2dFGRS at small scales do not show the acoustic peak. Moreover, we
can see a large scatter in the correlation function of the mocks,
with average values that do not follow the data (mocks show larger
correlations at intermediate scales and smaller at large scales).

Fig.~\ref{fig:pkall} shows the power spectrum calculated recently by
S\'anchez and Cole \cite{VMsanchez08} for the 2dFGRS and the
SDSS-DR5 survey. The expected acoustic oscillations are clearly
detected within the error bands.  These errors have been calculated
using mock catalogues generated from lognormal density fields with a
given theoretical power spectrum.

\subsection{Concluding remarks and challenges}

The expected value of the sound horizon at recombination
(Eq.~\ref{eq:soundh}) determined from the CMB observations can be
compared with the observed BAO scale in the radial direction at a
given redshift to estimate the variation of the Hubble parameter
with redshift $H(z)$. High accurate redshifts are needed to carry on
this test. Likewise, the BAO scale observed in redshift surveys
compared with its expected value provides us with a way to measure
the angular diameter distance, as a function of redshift $D_A(z)$.
As Nichols \cite{VMnichol07} points out this is similar, in a sense,
to the measurement of the correlation function in the parallel and
perpendicular directions to the line of sight, $\xi(\pi, r_p)$,
explained in Sec. 3.1.

There are several ongoing observational projects that will map a
volume large enough to accurately measure BAOs in the galaxy
distribution, some of them making use of spectroscopic redshifts
(i.e., AAT WiggleZ, SDSS BOSS, HETDEX, and WFMOS) and others making
use of photometric redshifts (i.e., DES, LSST, Pan-STARRS, and PAU),
all of them surveying large areas of the sky and encompassing
volumes of several Gpc$^3$. For an updated review see
\cite{VMfrieman08}. To deal with the uncertainties of the BAO
measurement due to different effects (non-linear gravitational
evolution, biasing of galaxies with respect to dark matter, redshift
distortions, etc.) is not easy, and accurate cosmological
simulations are required for this purpose.

The correlation function can be generalized to higher order (see the
contribution by Istvan Szapudi in this volume): the $N$-point
correlation functions. This allows to statistically characterize the
galaxy distribution with a hierarchy of quantities which
progressively provide us with more and more information about the
clustering of the point process. These measures, however, had been
difficult to derive with reliability from the scarcely populated
galaxy catalogs. The new generation of surveys will surely overcome
this problem.

There are, nevertheless, other clustering measures which provide
complementary information to the second-order quantities described
above. For example, the topology of the galaxy distribution measured
by the genus statistic provides information about the connectivity
of the large-scale structure. The topological genus of a surface is
the number of holes minus the number of isolated regions plus 1.
This quantity is calculated for the isodensity surfaces of the
smoothed data corresponding to a given density threshold (excursion
sets). The genus can be considered as one of the four Minkowski
functionals used commonly in stochastic geometry to study the shape
and connectivity of union of convex three-dimensional bodies. In 3-D
there are four functionals: the volume, the surface area, the
integral mean curvature, and the Euler-Poincar\'e characteristic,
related with the genus of the boundary (see the contribution by Enn
Saar in this volume).

The use of wavelets and related integral transforms is an extremely
promising tool in the clustering analysis of 3-D catalogs. Some of
these techniques are introduced in the contributions by Bernard
Jones, Enn Saar and Belen Barreiro in this volume.

\vskip 1.true cm

{\noindent{\bf Acknowledgements}}\\
I thank Enn Saar, Carlos Pe\~na and Pablo Arnalte for useful
comments and suggestions on the manuscript, Gert H\"utsi for the
data for Fig.~\ref{fig:peakxi}, Fernando Ballesteros and
Silvestre Paredes for help with the figures, and Pablo de la Cruz,
and Mar\'{\i}a Jes\'us Pons-Border\'{\i}a for allowing me to use
unpublished common work on the analysis of the large-scale
correlation function of the 2dFGRS in this review.  I acknowledge
financial support from the Spanish Ministerio de Educaci\'on y
Ciencia project AYA2006-14056 (including FEDER) and the Acci\'on
Complementaria AYA2004-20067-E.

\end{document}